\documentclass[conference]{IEEEtran}
\IEEEoverridecommandlockouts
\usepackage{geometry}
\usepackage{comment}
\usepackage{cite}
\usepackage{amsmath,amssymb,amsfonts}
\usepackage{algorithmic}
\usepackage{graphicx}
\usepackage{textcomp}
\usepackage{xcolor}
\usepackage{float}
\def\BibTeX{{\rm B\kern-.05em{\sc i\kern-.025em b}\kern-.08em
    T\kern-.1667em\lower.7ex\hbox{E}\kern-.125emX}}
\begin{document}

\title{Dynamic Ready Queue Based Process Priority Scheduling Algorithm
}

\author{\IEEEauthorblockN{Raghav Dalmia, Aryaman Sinha, Ruchi Verma, P. K. Gupta}
\IEEEauthorblockA{Department of Computer Science \& Engineering \\}
\IEEEauthorblockA{Jaypee University of Information Technology, Solan, 173 234, India\\
Email: {\{raghudalmia1712, aryan040501, ruchiverma612\}@gmail.com}, pkgupta@ieee.org,\\ }}

\maketitle

\begin{abstract}
CPU scheduling is the reason behind the performance of multiprocessing and in time-shared operating systems. Different scheduling criteria are used to evaluate Central Processing Unit Scheduling algorithms which are based on different properties of the system. Round Robin is known to be the most recurrent pre-emptive algorithm used in an environment where processes are allotted a unit of time and multiprocessing operating systems. In this paper, a reformed variation of the Round Robin algorithm has been introduced to minimise the completion time, turnaround time, waiting time and number of context switches that results in the better performance of the system. The proposed work consists of calculation of priority on the basis of the difference between time spent in ready upto the moment and arrival time of the process, to ease up the burden on the ready queue. We have also evaluated the performance of the proposed approach on different datasets and measured the different scheduling criteria
\end{abstract}

\begin{IEEEkeywords}
Burst Time, Arrival Time, CPU Scheduling, Time Quantum, Priority, Context Switches
\end{IEEEkeywords}

\section{Introduction}
The allocation and deallocation of system resources is performed by portion of the operating system known as scheduler and this mechanism is referred as CPU Scheduling \cite{b1}. Priority of the process is identified by a Scheduler, among multiple executable processes which are waiting for allocation of CPU for execution. CPU Scheduling are of two types, mainly, non pre-emptive, or pre-emptive. There exists various CPU scheduling algorithms like First Come First Serve (FCFS), Shortest Job First (SJF) as non pre-emptive and Shortest Remaining Time First (SRTF), priority scheduling as pre-emptive algorithm. 

In Priority Scheduling, processes with higher priority are carried out first, the process with lower one is paused and the higher one is completely executed \cite{b2}. Further, Round Robin is the most common pre-emptive Scheduling Algorithm, which is widely used due to its simple proof of work that provides equal chance to all processes and the expansive theory which can be further modified according to the user’s need, and not being much competitive \cite{b3}\cite{b4}. Here, RR is much simpler to implement and has much lower runtime overhead \cite{b5}. The efficiency of the Round Robin varies on the time quantum i.e., while the time quantum is small, a greater number of context switches takes place, whereas a large time quantum makes, the algorithm to work like FCFS causing starvation occurring to processes \cite{b6}.

In this paper, we have devised an algorithm, which mainly focusses on the reduction of different scheduling criteria, such as average turnaround time, average waiting time and number of context switches. Median of all the burst times is used as the time quantum for fulfilling the dynamic needs of the algorithm. Even the problem of priority which arises due to the large amount of time spent in ready queue is solved with the priority method of the algorithm. It is based on the difference of arrival time and time spent by the process in the ready queue. A threshold condition is also implemented to check whether the process is on the verge of completion and exit the queue. Terminating shorter processes early in the cycle to avoid the problem of starvation.

Many objectives are needed to be fulfilled for the better implementation of the algorithm. These objectives determine the overall performance of the algorithm over existing algorithms.
\begin{itemize}
\item  Reduces waiting time, which is the total time a process was in ready queue.
\item Reduces turnaround time, which is the total time between the arrival and completion of a process.
\item Reduces number of context switches, which refers to transition from one process to another due to expiring of time quantum, the process got interrupted, etc.
\item Overheads of added context switches which results from choosing small time quantum, and starvation which results from choosing large time quantum should be exempted.
\end{itemize}
Scheduling of processes is one of the important tasks. There have been many algorithms over the course of time, but Round Robin is the most simple and effective of all. The efficiency of algorithm based on RR depends on time quantum, whereas the accurate time quantum is which minimizes the number of context switches and avoids starvation. The proposed algorithm focussed on achieving many objectives for improved results over the existing algorithms. The performance of the algorithm largely varies upon different scheduling states which inspects the algorithm in different criteria. Fig. 1 represents the process of scheduling. 

\begin{figure}[htbp]
\centerline{\includegraphics[width=0.5\textwidth]{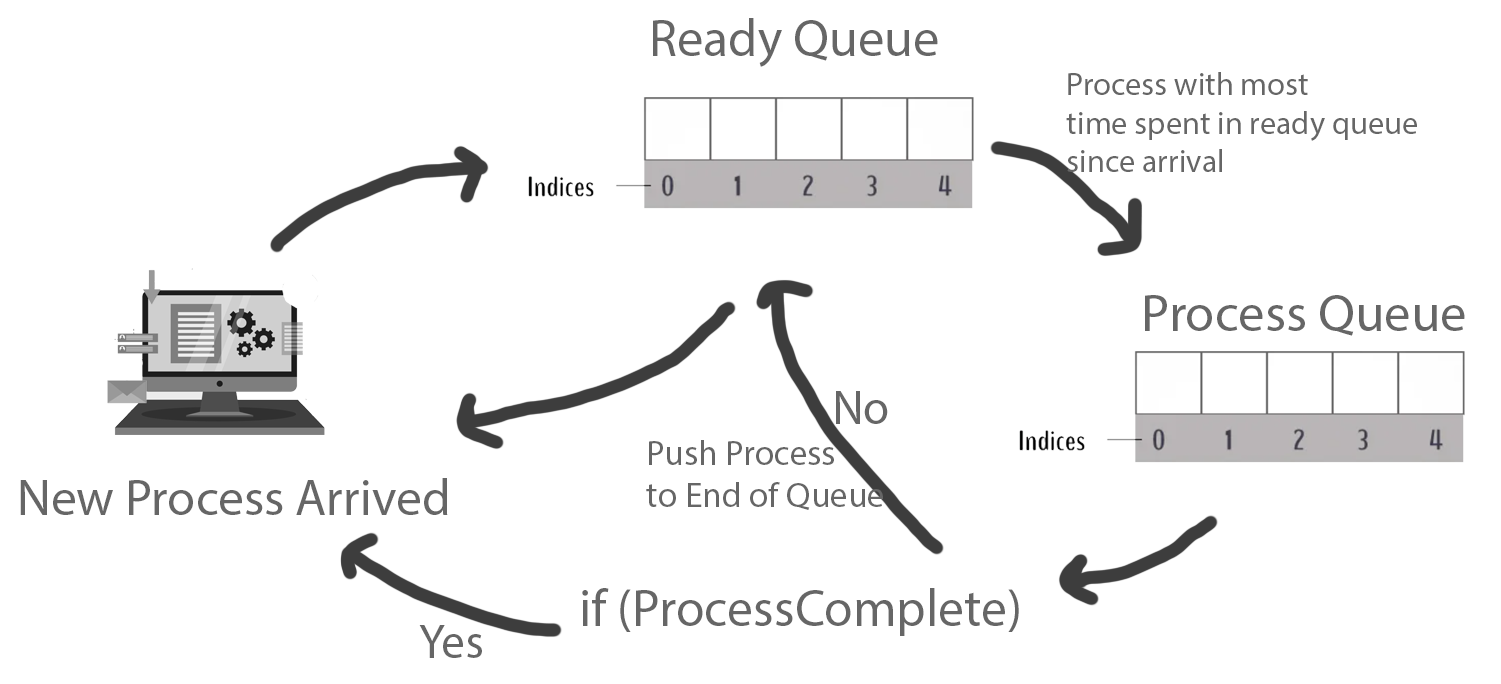}}
\caption{Representation of Scheduling Process.}
\label{fig1}
\end{figure}

\section{Related Work}
SRR Algorithm is widely administered in majority of the OS for better CPU performance. The search of an algorithm which improves and optimizes the performance of CPU by maximizing utilization and minimizing average waiting time and turnaround time, led to the proposal of many variants of SRR.

\subsection{Dynamic Time Quantum techniques}

In \cite{b1}, Mostafa and Amano have proposed a version of Standard Round Robin (SRR) by using clustering techniques to group similar processes in clusters. The clustering metrics are Burst Time (BT), Process Weight (PW), and Number of Context Switches (NCS). In subsequent rounds, the time slice of each cluster is calculated using the weight of each cluster with a threshold condition in consideration. If a new process arrives, it is added to the end of the queue. In \cite{b10}, Tithi et al. have implemented a dynamic version of SRR which focuses on progressive time quantum which needs to be consistently adjusted according to the remaining burst time of currently running processes. Like Lipika's and Sulalah’s \cite{b7} \cite{b11} algorithm, processes are sorted in ascending order in this algorithm too. In \cite{b8} Alsheikhy et al. have proposed a dynamic way to keep the time quantum comparatively smaller than the value where the algorithm acts as FCFS by computing average of two highest burst times and later the average of two lowest arrival times which were estimated value for single use only; later, the average of arrival time for only lowest process are subtracted. In \cite{b12}, Tajwar and  Nuruddin  have presented a version of Round Robin in which time quantum is dynamically adjusted in the starting of subsequent rounds which is equal to the mean of remaining burst times of processes in every round. In \cite{b13}, Samir and Shahenda  have proposed a SRR and SJF based mix scheduling algorithm known as SJF and RR based dynamic time quantum (SRDQ). In their approach, two separate queues Q1 and Q2 have been divided from the major ready queue; for processes shorter than the median and for processes longer than the median, respectively. Further, in subsequent rounds allocations of new time quantum is done depending on the burst time and median to the processes. This algorithm is a combination based algorithm with a basis of dynamic time quantum.

\subsection{Threshold based techniques}
In \cite{b11}, Samkit and Sulalah have presented a version of SRR where “Smart Time Quantum” and “Delta” are introduced and computed. These criteria let processes with less burst time to terminate rather than to wait for an entire round to execute. Thus, these two provide flexibility to the processes and reduce waiting time by providing extra CPU time. In \cite{b9}, Samih and Hirofumi have suggested a better version of SRR to tackle the low-scheduling cost and short processes known as Adjustable Round Robin (ARR). A threshold condition is checked for determination of interruption of the running process due to completion of its time slice considered under SRR or execution until termination.

\subsection{Priority based techniques}

In \cite{b7}, Lipika came up with dynamic scheduled round robin algorithm by reinitialising the time quantum at the beginning of every round. The time slice is calculated using the remaining burst times in the following rounds. Precedence factor is the ratio of the remaining CPU burst of a process and the relative waiting time which is calculated for every process in the starting of each round. It is therefore, used in determining the priority order of processes for that particular round. In \cite{b14}, Pawan et al. have proposed a variant of Pre-emptive scheduling to avoid starvation by prioritising processes with short burst time or remaining burst time over processes with long burst time. Initially priority is given according to the inverse of burst time. Afterwards priority is directly proportional to waiting time and inversely proportional to remaining burst time.

\section{Problem Formulation}
In FCFS, there arises the problem of starvation due to the first come first serve nature of the priority queue as for the completion of a process, it runs for the complete burst time till that time all the processes wait in the queue, whereas, in SRR, the process can only run for a number of time quantum till a process with higher priority arrives, even time quantum cannot adjust according to the processes. Therefore, when the time quantum is too small, then context switches count increases which lead to starvation and if it is too long, the algorithm works like FCFS.

If we have n processes

The completion time for the $n^{th}$  process in FCFS = $\sum_{n=1}^{n} (BT_{n})$

whereas that in Round Robin = $\sum_{n=1}^{n-1} (TQ)$

and,

Turnaround time for $n^{th}$ process in FCFS = $\sum_{n=1}^{n} [(BT_{n})-(AT_{n})]$ 

whereas in Round Robin = X$\sum_{n=1}^{n-1} [(y_{n})-(AT_{n})]$

where X is dynamic number of rounds and $y_{n} = TQ_{n} \forall TQ$ if $RBT_{n}>TQ$ or $TQ_{n} = RBT_{n}$ if $RBT_{n}<TQ$

It is analysed that FCFS is poor in terms of performance criteria, whereas RR performs better when there is accurate time quantum. Accurate time quantum can result in reduced completion time and turnaround time.

\section{Proposed Approach}
The proposed work is an advanced interpretation of the RR scheduling algorithm. Focusing on enhancement of time quantum and priority of processes entering the ready queue leads to better scheduling states, which are calculated in the following subsection. The proposed technique consists of two steps: Calculation of Dynamic Time Quantum and finally a priority order for the processes.

\subsection{Calculation of Dynamic Time Quantum}
Time Quantum for the $i^{th}$ round, $TQ_{i}$, is calculated from \eqref{eq1}:
\begin{equation}
TQ_{i}=Med(BT_{n})=\begin{cases}
			BT \left[ \frac{n}{2} \right], & \text{if $n$ even}\\
            BT \left[n + \frac{1}{2} \right], & \text{if $n$ odd}
		 \end{cases}\label{eq1}
\end{equation}

where $BT_{n}$ is the burst time of the $n^{th}$ process, and [X] is the greatest integer function which represents the largest integer value smaller than or equal to X. Median is preferred over any other measure of central tendency, as the value of median does not depend on all the values of the dataset. Thus, when the data is skewed, the effect on the median is smaller as it is not affected by unsymmetrical values.

\subsection{Calculation of Priority Order}
In the second step, priority order of processes, the time spent in ready queue by a process, $TRQ_{n}$, is calculated from \eqref{eq2}:
\begin{equation}
TRQ_{n}=\sum_{i=1}^{i-1} (TQ)_{i}(k-1)_{i}\label{eq2}
\end{equation}

where, $TQ_{i}$ is the Time Quantum for $i^{th}$ round, and $k_{i}$ is equivalent to number of processes in ready queue in $i^{th}$ round. The priority to processes in the $i^{th}$ round, $P_{i}$, is calculated using \eqref{eq3}:
\begin{equation}
P_{i}=\begin{cases}
			\uparrow AT_{n}, & \text{if $i$=1 \& $k$}<{n/2}\\
            \uparrow \left[TRQ_{n}-AT_{n} \right], & \text{if $i$}\neq{1}
		 \end{cases}\label{eq3}
\end{equation}

where, $AT_{n}$ and $TRQ_{n}$ are the Arrival Time and Time spent by the $n^{th}$ process in the ready queue, respectively. Priority order of processes is reinitialized at the beginning of every round for the exit of smaller processes accordingly.

Each process in the queue executes for $TQ_{i}$. It is dynamically calculated at the beginning of each cycle using (1), on the basis of remaining burst time till the current round. Moreover, there is an opportunity for the processes which are on the verge of completion to get advanced in the queue to conclude and exit the queue. Threshold value is computed to permit the processes with remaining burst time equal to or less than 4\% of the original burst time to resume execution and terminate. As a result, the number of processes will reduce within the ready queue with the termination of short processes fairly faster, which might result in comparatively less average TAT and WT. The remaining burst time of the $n^{th}$ process, $RBT_{n}$, is computed by \eqref{eq4}:
\begin{equation}
RBT_{n}=BT_{n}-Med(BT_{n})\label{eq4}
\end{equation}

When $RBT_{n}$ is equal to zero or the threshold condition is satisfied, the process exits the queue. Arrival of new processes is marked by them being placed at the tail of the queue to be scheduled and execute in subsequent rounds. Here, Fig. 2 represents the flowchart of the proposed algorithm.
\begin{figure}[H]
\centerline{\includegraphics[width=0.45\textwidth, height=0.8\textwidth]{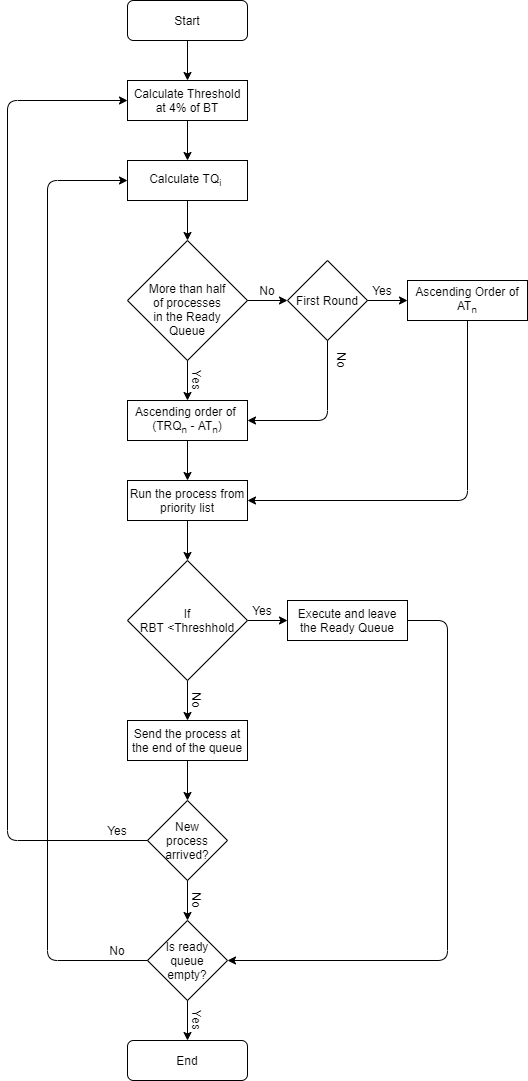}}
\caption{Representation of Scheduling Process.}
\label{fig2}
\end{figure}

\subsection{Illustration}
For understanding the better functioning of the proposed algorithm, let’s consider and example as per Table 1. Further, to demonstrate the concept consider the dataset used in the experimental implementation and consist of 6 processes with constant time quantum of 3tu.
\begin{table}[htbp]
\caption{Dataset}
\begin{center}
\begin{tabular}{|c|c|c|}
\hline
\textbf{\textit{ID}}& \textbf{\textit{Arrival Time}}& \textbf{\textit{Burst Time}} \\
\hline
P1 & 5 & 5 \\ \hline
P2 & 4 & 6 \\ \hline
P3 & 3 & 7 \\ \hline
P4 & 1 & 9 \\ \hline
P5 & 2 & 2 \\ \hline
P6 & 6 & 3 \\ \hline
\end{tabular}
\label{tab1}
\end{center}
\end{table}

The Gantt chart as shown in Fig. 3, represents when the processes are scheduled according to the SRR.

\begin{figure}[htbp]
\centerline{\includegraphics[width=0.5\textwidth]{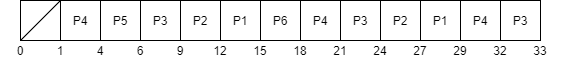}}
\caption{Representation of SRR \cite{b2}.}
\label{chart1}
\end{figure}

The waiting time for the 6 processes are 19tu, 17tu, 23tu, 22tu, 2tu and 9tu, respectively. Thus, resulting in average waiting time of 15.34tu, and average turnaround time of 20.67tu. On the contrast, suppose TQ and Priority are calculated according to \eqref{eq1} and \eqref{eq3} mentioned.
\begin{table}[htbp]
\caption{Dynamic Time Quantum for each subsequent cycle}
\begin{center}
\begin{tabular}{|c|c|c|}
\hline
& \textbf{\textit{CYCLE 1}}& \textbf{\textit{CYCLE 2}} \\
\cline{2-3} 
\textbf{ID} & \textbf{\textit{TQ = 6tu}}& \textbf{\textit{TQ = 3tu}} \\
\cline{2-3} 
& \textbf{\textit{BT}}& \textbf{\textit{RBT}} \\
\hline
P1 & 5 & - \\ \hline
P2 & 6 & - \\ \hline
P3 & 7 & 1 \\ \hline
P4 & 9 & 3 \\ \hline
P5 & 2 & - \\ \hline
P6 & 1 & - \\ \hline
\end{tabular}
\label{tab2}
\end{center}
\end{table}

The Gantt chart as shown in Fig. 4 has been prepared for the proposed algorithm as per following:
\begin{figure}[htbp]
\centerline{\includegraphics[width=0.5\textwidth]{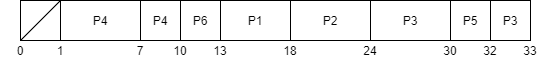}}
\caption{Representation of Proposed Algorithm.}
\label{chart2}
\end{figure}

The waiting time for the 6 processes are 7tu, 14tu, 23tu, 0tu, 30tu and 10tu, respectively. Thus, resulting in average waiting time of 12.84tu, and average turnaround time of 18.17tu. The subsequent cycles are shown in Table 2.

\section{Experimentation}
\begin{table}[htbp]
\caption{Testbed for Performance Evaluation}
\begin{center}
\begin{tabular}{|c|c|}
\hline
Processor & Intel core i5-8250U (1.60 GHz) \\ \hline
RAM & 8GB \\ \hline
Memory & 256GB SSD, 2TB HDD \\ \hline
Operating System & Windows 10 \\ \hline
Simulation & Python 3.9.7 (Google Colab) \\ \hline
\end{tabular}
\label{tab3}
\end{center}
\end{table}

\subsection{Dataset}
Simulated dataset as shown in Table 4 is used for the experimental purposes and to test the performance of proposed algorithm. Even for the one with which comparison is to be made. Each dataset contains randomly generated processes with separate arrival time and burst time. Therefore, the dataset distinguish all the factors.

\begin{table}[!ht]
\caption{Dataset vs. Number of Processes.}
\begin{center}
\begin{tabular}{|c|c|}
\hline
\textbf{\textit{Dataset ID}}& \textbf{\textit{Number of Processes}} \\
\hline
1 & 4 \\ \hline
2 & 5 \\ \hline
3 & 5 \\ \hline
4 & 6 \\ \hline
5 & 6 \\ \hline
6 & 10 \\ \hline
7 & 10 \\ \hline
8 & 15 \\ \hline
9 & 15 \\ \hline
10 & 20 \\ \hline
\end{tabular}
\label{tab4}
\end{center}
\end{table}

\begin{table*}[!ht]
\caption{Percentage Evaluation of the Proposed Algorithm with  SRR \cite{b2}}
\begin{center}
\begin{tabular}{|c|c|c|c|c|c|c|c|c|c|}
\hline
\textbf{Dataset}&\multicolumn{3}{|c|}{\textbf{Proposed}}&\multicolumn{3}{|c|}{\textbf{SRR}}&\multicolumn{3}{|c|}{\textbf{\% Improvement}} \\
\cline{2-10} 
\textbf{ID}& \textbf{\textit{TAT}}& \textbf{\textit{WT}}& \textbf{\textit{NCS}}& \textbf{\textit{TAT}}& \textbf{\textit{WT}}& \textbf{\textit{NCS}}& \textbf{\textit{TAT}}& \textbf{\textit{WT}}& \textbf{\textit{NCS}} \\
\hline
1& 6& 3& 4& 7.25& 4.25& 6& 17.24& 29.41& 33.33 \\ \hline
2& 8.2& 4& 6& 9.4& 5.2& 10& 10.63& 23.07& 40\\ \hline
3& 14& 7.4& 6& 18.4& 11.8& 17& 23.91& 37.16& 64.7\\ \hline
4& 41.5& 28.58& 7& 42.33& 29.41& 12& 1.96& 2.83& 41.66\\ \hline
5& 18.17& 12.84& 9& 20.67& 15.34& 13& 14.86& 19.81& 46.15\\ \hline
6& 493.3& 353.1& 17& 664.5& 529.5& 72& 25.76& 33.31& 76.38\\ \hline
7& 235.7& 196& 15& 275.8& 236.1& 50& 14.53& 16.98& 70\\ \hline
8& 1929.67& 1706.47& 26& 2424.2& 2204& 167& 20.39& 22.57& 84.43\\ \hline
9& 2076.2& 1827.87& 23& 2799.6& 5249.6& 187& 25.84& 28.3& 89.61\\ \hline
10& 1951.6& 1758.05& 3& 2908.75& 2699.5& 165& 32.9& 34.48& 81.21\\ \hline
\hline
Average& 677.43& 589.73& 14.4& 917.09& 828.47& 69.9& 18.80& 24.79& 62.74\\ \hline
\hline
\end{tabular}
\label{tab5}
\end{center}
\end{table*}

\subsection{Performance Evaluation}
The algorithm is performed using the multiple set of randomly generated dataset on a system with specifications mentioned in Table 3. The authors compared the proposed algorithm with SRR for the compilations of accurate results. For the efficiency of results, dataset with varying, number of processes, arrival time and burst time are used. The resulting criteria depends upon the number of processes, therefore, number of processes vary from as small as 4 to as large as 20, so that there is no discrepancy in results. More number of processes in ready queue results in increase in total runtime of processes thus, affecting the average turnaround and waiting time in a negative way. Runtime is also affected by long burst times and delayed arrival of processes. Here, Table 5 represents the comparison of the proposed approach and the Standard Round Robin in terms of Average TAT, Average WT and NCS. Fig. 5 and Fig. 6 are pointing towards the supremacy for proposed approach over Standard Round Robin. Whereas, Fig. 7 manifest the percentage improvement in the all the scheduling states in every dataset.

\begin{figure}[!ht]
\centerline{\includegraphics[width=0.5\textwidth, height=0.4\textwidth]{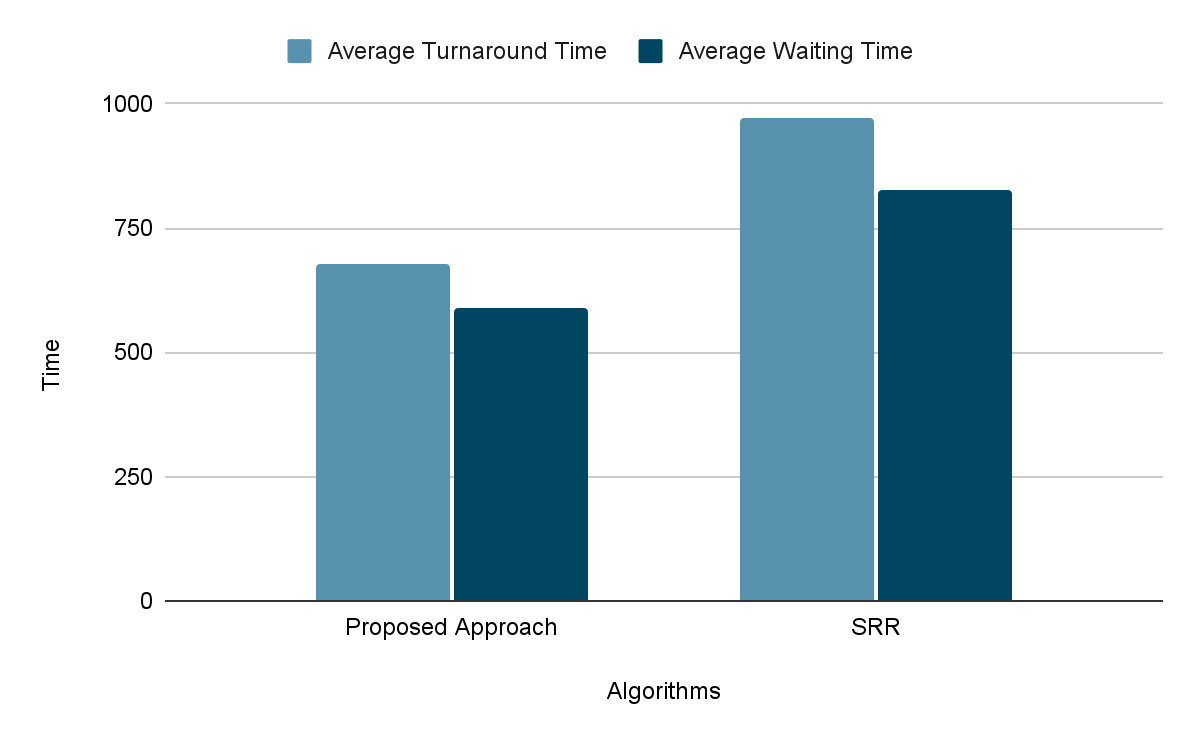}}
\caption{Comparison of Algorithms on the basis of Average Turnaround Time and Average Waiting Time.}
\label{fig3}
\end{figure}

\begin{figure}[!ht]
\centerline{\includegraphics[width=0.5\textwidth, height=0.5\textwidth]{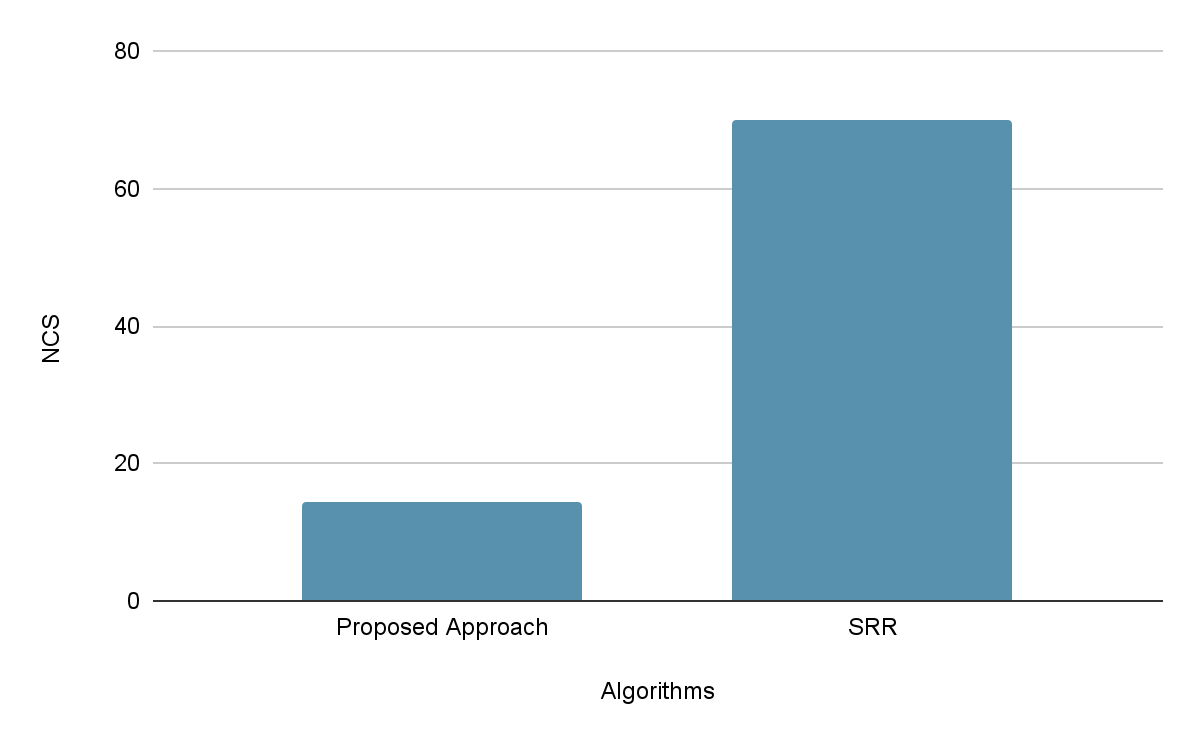}}
\caption{Comparing Algorithm’s NCS.}
\label{fig4}
\end{figure}

\begin{figure}[!ht]
\centerline{\includegraphics[width=0.5\textwidth, height=0.5\textwidth]{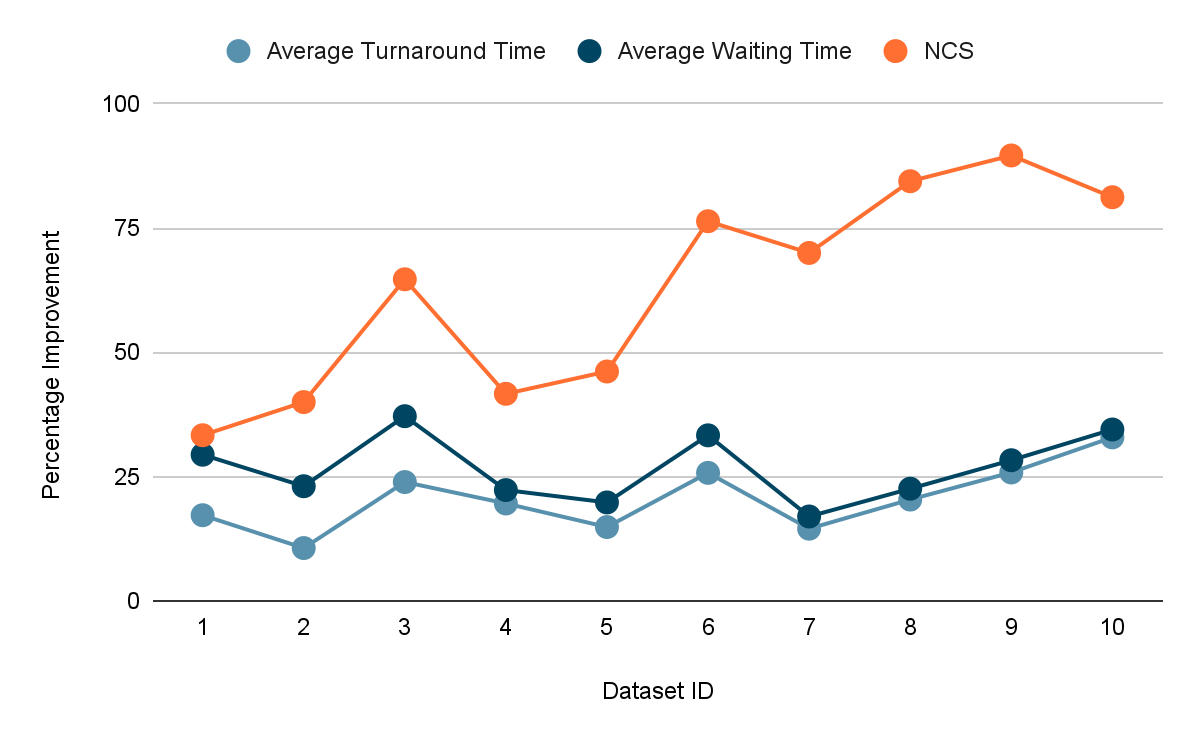}}
\caption{Percentage Improvement of different Scheduling States.}
\label{fig5}
\end{figure}

\section{Conclusion}
The paper proposes a variant of Dynamic Round Robin for multiprocessing and time-sharing machines. Reduction of scheduling duration and increase the processing speed is the main purpose of these machines. This algorithm works on dynamic time quantum, as compared to SRR which has a fixed time quantum till the scheduling continues. The proposed algorithm benefits from the method of calculating time quantum and calculating priority for the processes. The features are updated after every subsequent round till all the processes terminate and exit the queue. The processes which are close to termination are even given a chance to exit in that particular round instead of the next round, this in turn reduces the processes in the queue thus, affecting the scheduling states.

\end{document}